\documentclass[useAMS,usenatbib]{mn2e}
\usepackage{color}
\usepackage{epsfig,amsmath,natbib}
\def\lsim{\;\raise0.3ex\hbox{$<$\kern-0.75em\raise-1.1ex\hbox{$\sim$}}\;}
\def\gsim{\;\raise0.3ex\hbox{$>$\kern-0.75em\raise-1.1ex\hbox{$\sim$}}\;}

\newcommand{\Lmin}{L_\mathrm{min}}
\newcommand{\Lmax}{L_\mathrm{max}}
\newcommand{\Kmin}{k_\mathrm{min}}
\newcommand{\Kmax}{k_\mathrm{max}}

\def\cmc{\rm ~cm^{-3}}
\def\kms{\rm ~km~s^{-1}}

\def\cmc{\rm ~cm^{-3}}

\def \kms {\rm ~km~s^{-1}}

\def\enf{\rm ~erg~cm^{-2}~s^{-1}}
\def\arcmin{\hbox{$^\prime$}}
\def\arcsec{\hbox{$^{\prime\prime}$}}

\title{A Model of Polarized X-ray Emission from Twinkling Synchrotron Supernova Shells}
\author[A.M.Bykov, Yu.A.Uvarov, J.B.G.M.Bloemen, J.W. den Herder, J.S.Kaastra]{A.M.Bykov$^{1}$\thanks{E-mail:byk@astro.ioffe.ru}, Yu.A.Uvarov$^{1}$, J.B.G.M.Bloemen$^{2}$, J.W. den Herder$^{2}$ and
J.S.Kaastra$^{2}$\\
$^{1}$Ioffe Institute for Physics and Technology, 194021
St.Petersburg, Russia\\
$^{2}$SRON Netherlands Institute for Space Research, Utrecht, The
Netherlands}

\begin{document}

\date{Accepted 2 July 2009. Received 2009 July 2; in original form 2009 May 2.}

\pagerange{\pageref{firstpage}--\pageref{lastpage}} \pubyear{2002}

\maketitle

\label{firstpage}

\begin{abstract}
Synchrotron X-ray emission components were recently detected in many
young supernova remnants (SNRs). There is even an emerging class -
SN1006, RXJ1713.72-3946, Vela Jr, and others - that is {\sl
dominated} by non-thermal emission in X-rays, also probably of
synchrotron origin. Such emission results from electrons/positrons
accelerated well above TeV energies in the spectral cut-off regime.
In the case of diffusive shock acceleration, which is the most
promising acceleration mechanism in SNRs, very strong magnetic
fluctuations with amplitudes well above the mean magnetic field must
be present. Starting from such a fluctuating field, we have
simulated images of {\sl polarized} X-ray emission of SNR shells and
show that these are highly clumpy with high polarizations up to
50\%. Another distinct characteristic of this emission is the strong
intermittency, resulting from the fluctuating field amplifications.
The details of this "twinkling" polarized X-ray emission of SNRs
depend strongly on the magnetic-field fluctuation spectra, providing
a potentially sensitive diagnostic tool. We demonstrate that the
predicted characteristics can be studied with instruments that are
currently being considered. These can give unique information on
magnetic-field characteristics and  high-energy particle
acceleration in SNRs.
\end{abstract}

\section{Introduction}
Electrons and positrons accelerated to TeV energies by diffusive
shock acceleration (DSA) in SNR shells will efficiently radiate in
X-rays in the associated magnetic fields \citep[e.g.,][]{rs81}. A
recent review on `SNRs at high energy' is given by
\citet{reynolds08}. In some sources (e.g. SN1006, RXJ1713.72-3946
and Vela Jr), the synchrotron component is dominating the X-ray
emission, whereas in others such as Cas~A it is not easy to
distinguish the synchrotron component from the bremsstrahlung
emission. Mapping of the {\sl polarized} X-ray emission from SNRs
would allow to separate out and study the synchrotron components.

With the high-resolution imaging capability of {\sl Chandra}, likely
synchrotron structures are already seen in the X-ray images of
various SNRs \citep[e.g.,][]{vl03,bambaea06,pf08}. The observed
non-thermal emission is concentrated in very thin (arcseconds width)
filaments and clumps and has typically a rather steep spectrum with
an exponential roll-off. In addition, \citet{uchiyamaea07} reported
variability of such X-ray hot spots in the shell of SNR
RXJ1713.72-3946 on about a one-year timescale. The thin filamentary
structures can be naturally explained in the DSA scenario: options
are 1) a narrow spatial extend of the TeV-regime electron population
caused by efficient electron cooling due to synchrotron energy
losses in the vicinity of the SNR shock with strong magnetic-field
amplification \citep[e.g.,][]{vl03,bambaea05,vink08} and 2) the
observed narrow filaments are limited by magnetic field damping and
not by the energy losses of the radiating electrons
\citep[e.g.,][]{pyl05}.

Polarized X-ray emission - {\sl from any source} -
was observed so far only in  very few
cases. Observations of the Crab Nebula with X-ray polarimeters aboard
{\sl OSO-8} \citep{novickea72,weisskopfea76} revealed a polarized flux
of about 15\% at few keV energies (also detected with the {\sl IBIS}
detector on {\sl INTEGRAL} by \citet{forotea08}).
Recently,  \citet{gotzea09} reported variable
polarized emission at 200-800 keV from
GRB 041219A with {\sl IBIS}. Very little else can be reported thus far.

Efficient DSA of protons and electrons in supernova shells requires
turbulent magnetic fields, with energy densities that are a
substantial fraction of the shock ram pressure \citep[e.g.,][]{
be87, md01, hillas05, bell04, ab06, veb06}. Both regular and
stochastic magnetic fields determine the spectra and maps of
synchrotron radiation of high-energy electrons and positrons from
SNRs.

A model of non-thermal {\sl radio} emission from SNRs, accounting
for the orientation of the regular ambient magnetic field, was
presented recently by \citet{petrukea09}. These authors synthesized
radio maps of SNRs, making various assumptions on the dependence of
the electron injection efficiency on the shock obliquity. Their
method uses the azimuthal profile of the radio surface brightness as
a probe of the orientation of the ambient magnetic field. The effect
of random magnetic fields in supernova shells on radio synchrotron
emission was addressed by \citet{sp09}. They discussed the emission
and transport of polarized radio-band synchrotron radiation near the
forward shocks of young shell-type supernova remnants with a strong
amplification of the turbulent magnetic field. Modeling the magnetic
turbulence was done as a superposition of waves at a particular
moment in time; no time evolution was considered. They found that
isotropic strong turbulence produces only weakly polarized radio
emission even in the absence of internal Faraday rotation. If
anisotropy is imposed on the magnetic-field structure, then the
degree of polarization can be significantly increased, if the
internal Faraday rotation is inefficient.

It has long been known that random directions of magnetic fields in
addition to Faraday rotation may strongly reduce the {\sl average}
polarization of synchrotron emission sources
\citep[e.g.,][]{westfold59, cs86,sp09}. This explains the relatively
low polarization frequently observed for radio synchrotron sources.
However, as we will show below, the turbulent magnetic fields that
reduce the {\sl average} polarization can result in highly polarized
patchy structures potentially observable in high resolution images
at X-rays.

\citet{reynolds98}
simulated X-ray synchrotron images assuming a regular magnetic field
and distributions of ultra-relativistic electrons accelerated by a
forward shock using age-limited and loss-limited parameterizations.

The effect of turbulent magnetic fluctuations (including field {\sl
magnitude} fluctuations) on synchrotron X-ray emission images was
recently addressed by \citet{bue08}. A system of finite size filled
with a random magnetic field was modeled and used to construct
synchrotron emission maps of a source with kinetically simulated
distributions of ultra-relativistic electrons. The random field was
composed of a superposition of magnetic fluctuations (transverse
plane waves propagating with some phase velocity) with random phases
and a given spectrum of amplitudes. Accounting for the field {\sl
magnitude} fluctuations was especially important in view of the
dependence of the emissivity on the local magnetic field (further
addressed below). A particulary strong dependence occurs in the
cut-off regime of the synchrotron spectrum (also further addressed
below). \citet{bue08} found that non-steady structures (dots,
clumps, and filaments) typically arise, in which the magnetic field
reaches exceptionally high values. These magnetic-field
concentrations dominate the synchrotron maps, with an evolving,
intermittent, and clumpy appearance. The modeling showed that the
overall efficiency of synchrotron radiation from the cut-off regime
of the electron spectrum can be strongly enhanced in a turbulent
field with some $\sqrt{\langle B^2\rangle}$, compared to emission
from a uniform field of the same magnitude $\sqrt{\langle
B^2\rangle}$, but of just a random direction. Strong temporal
variations of the brightness of small structures were found, with
time scales much shorter than variations in the underlying particle
distribution. The variability time scale depends on the phase
velocity  and the spectrum of magnetic fluctuations. The simulated
structures indeed resemble the 'twinkling' structures that are
observed in X-ray images of some supernova remnants.

The same electrons that are producing X-ray synchrotron emission
will emit TeV photons by inverse-Compton scattering.
Both processes are of fundamental importance for our understanding
of high-energy particle acceleration and the distinction
between leptonic and nucleonic contributions to the observed
gamma-ray emission \citep[e.g.,][]{ahar_rxj1713, ahar_rcw86}.
Gamma-ray images of a SNR with efficient DSA in different circumstellar
environments were constructed by \citet{leeea08}.

In this paper we expand upon the work of \citet{bue08}, modeling
the specific features of the polarized
synchrotron emission arising from the stochastic nature of
magnetic fields of young SNR shells. In \S{\ref{model}} we describe
the simulation setup that includes the kinetic model of a TeV regime
electron distribution and a
simulation of a random magnetic field with different fluctuation spectra.
In \S{\ref{maps}} we present the resulting polarized
synchrotron emission maps for different X-ray energies and
different magnetic fluctuation spectra.
In \S{\ref{discussion}} we discuss the observational perspective.

\section{The model}\label{model}

In order to construct maps of polarized synchrotron emission from SNR
shells, it is convenient to use the local densities of the Stokes
parameters. Because of the additive property of the Stokes
parameters $\tilde{I}, \tilde{Q},\tilde{U}, \tilde{V}$ for
incoherent photons, we can integrate these over the line of sight
weighted with the distribution function of
radiating particles. The degree of
polarization is determined in a standard way as $\Pi = \sqrt{Q^2 +
U^2 + V^2}/I$.

The synchrotron emission is characterized by a coherence length
$l_f$ that is of the order of a MeV electron gyro-radius \citep[see
e.g.][]{rl79}. In the simulation we only consider the effects of
magnetic fluctuations having scales that are much larger than $l_f
\sim m_ec^2/e\sqrt{\langle B^2\rangle}$. This is because in the
nonlinear DSA modeling of non-relativistic SNR shocks the magnetic
fluctuation spectra are expected to fall down steeply at spatial
scales below the gyro-radius of a GeV proton \citep[see for instance
Fig.3 in][]{veb06}. That means that the fluctuation wavenumbers $k$
satisfy $k \cdot l_f \ll $ 1.
Therefore, neglecting the magnetic fluctuations of the scale less or
comparable to $l_f$, we apply the standard formulae  \citep[see
e.g.][]{gs65} for the synchrotron power of a single particle of
Lorentz factor $\gamma \gg 1$ in the simulated random magnetic field
composed of the long-wavelength MHD fluctuations. Then we integrate
this power over the line of sight through the system filled with
random field fluctuations.

The modeling of the polarized synchrotron emission from relativistic
shocks of GRBs, pulsar wind nebulae and AGNs objects would likely
require strong small scale magnetic fluctuations of wavenumbers $k
\cdot l_f \sim $ 1 and will be discussed elsewhere. The first
particle-in-cell simulations of relativistic shocks in unmagnetized
electron-positron pair plasmas \citep[see e.g.][]{spitkovsky08} have
demonstrated the feasibility of self-consistent modeling of pair
acceleration to energies above 100 times that of the thermal energy.
The simulated nonthermal particles were carrying about 10\% of the
downstream thermal energy, promising potential applications to the
modeling of polarized synchrotron emission from GRBs, blazars and
pulsar wind nebulae.

We start with the spectral flux densities
$p^{(1)}_{\nu}(\theta,\gamma)$ and $p^{(2)}_{\nu}(\theta,\gamma)$
with two principal directions of polarization radiated by a particle
with Lorentz factor $\gamma$, as given by \citet{gs65} [their
Eqs.(2.20)]. Here $\theta$ is the angle between the local magnetic
field ${\bf B}({\bf r}, t)$ and the direction to the observer. In
the case of a random magnetic field it is convenient to use the
local spectral densities of the Stokes parameters expressed through
$p^{(1)}_{\nu}$ and $p^{(2)}_{\nu}$:
\[
\hat{\tilde{S}} = \left(\begin{array}{c}\tilde{I}({\bf r},t, {\nu})\\ \tilde{Q}({\bf r}, t,{\nu})\\
\tilde{U}({\bf r},t, {\nu})\\ \tilde{V}({\bf r},t, {\nu})
\end{array} \right)= \left(\begin{array}{c}p^{(1)}_{\nu} + p^{(2)}_{\nu}\hspace*{1.2cm}\\
 (p^{(1)}_{\nu} - p^{(2)}_{\nu})\cdot \cos 2\chi\\
(p^{(1)}_{\nu} - p^{(2)}_{\nu})\cdot \sin 2\chi\\(p^{(1)}_{\nu} -
p^{(2)}_{\nu}) \cdot \tan 2\beta\end{array} \right)
\]
where the angle $\chi$ is between the major axis of the polarization
ellipse and a coordinate in the plane perpendicular to the observer
direction, and $\tan \beta$ is determined by the ratio of the minor
and major axes of the ellipse \citep{gs65}.

\begin{figure}
\centering {
 \rotatebox{0}{
{\includegraphics[height=5.6cm, width=7.5cm]{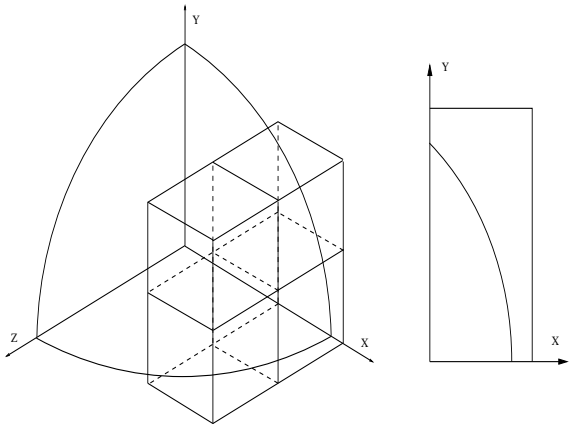} }}}
\caption{Geometry of the simulated supernova shell. The left panel
shows half of the shell quarter with the boxes in which the random
magnetic field was simulated. The local densities of the Stokes
parameters were then integrated over the line of sight along the
axis $-\infty < z < \infty$ (i.e. the two half-quarters of the
sphere). The right panel is the resulting projection that is shown
in the simulated maps.} \label{geom}
\end{figure}

\subsection{Random magnetic field and particle distribution}\label{field}
Synchrotron X-ray emission is radiated by 10 TeV regime electrons
since the magnetic field amplitude in SNR shells is typically below
a mG. Efficient DSA of high-energy particles requires a substantial
amplification of magnetic field fluctuations in the vicinity of the
shock; see e.g. \citet{bell78,be87,md01}. Magnetic-field
amplification mechanisms due to cosmic-ray instabilities in
nonlinear DSA were proposed recently by
\citet{bell04,ab06,veb06,plm06,vbe08,zp08}. The models predict
amplified magnetic-field amplitudes well above the interstellar
field in the far upstream of the shock. These current models are
suited to estimate the amplitudes and spectra of amplified magnetic
fluctuations averaged over some macroscopic spatial and temporal
scales. The averaged magnetic-field spectra available from the
models are appropriate to model energetic particle spectra, but do
not allow to simulate the synchrotron images of SNR shells and to
judge about their temporal evolution.

In reality  the distribution of the
emitting electrons is a random function of position and
particle energy because of the stochastic nature of both the
electromagnetic fields and the particle dynamics. However, no
self-consistent treatment of such a particle distribution in strong
magnetic turbulence is available. Rigorous modeling of the
magnetic-field structure and evolution should invoke
simultaneously fully
nonlinear PIC-type simulations of the collisionless shock,
supersonic flow, and the effect of the high-energy particles.
A microscopic selfconsistent description of magnetic-field
fluctuations that are strongly coupled with electric currents of
accelerated particles is not feasible yet for non-relativistic shock
simulations in SNRs \citep[see appendix in][for a
discussion]{vbe08}.

Therefore, to model the synchrotron SNR images we simulated local
statistically stationary random magnetic fields of given spectra
using the technique described in \citet{bue08}. The statistically
isotropic and homogeneous random field was constructed as a sum over
a large number of plane waves with wave vector, polarization, and
phase chosen randomly. In the simulation presented below we assume
a plane wave frequency $\omega_n({\bf k}_n) = v_{\rm ph}\cdot k_n$
parameterized with a phase velocity $v_{\rm ph}$. The spectral
energy density of the magnetic-field fluctuations of wavenumber
$k$ is described as $W(k) \propto k^{-\delta}$, where $\delta$ is
the spectral index.
\[
\langle B^2\rangle= \int_{\Kmin}^{\Kmax} dk\, W(k),
\]
The average square magnetic field $\langle B^2\rangle$, the spectral
index $\delta$, and the wavenumber range ($\Kmin$, $\Kmax$) are the
input parameters of the model. The spectral index $\delta$ in the
standard DSA scenario is expected to be in the range $1 \leq \delta
\leq 2$.

Then the local spectral emissivity of polarized synchrotron emission
was determined at various times using a calculated model
distribution of electrons (or positrons). The kinetic model used to
simulate the electron distribution was described in detail by
\citet{bceu00}. The spatially inhomogeneous electron distribution
function is calculated from the kinetic equation for electrons at a
SNR shock that uses piece-wise parametrization of the particle
diffusion coefficient to account for both Fermi I and II types
accelerations and that is consistent with the magnetic fluctuation
spectrum. The model assumes a diffusion coefficient $\kappa(p)
\propto p^a$, where $a = 1$  for TeV-regime electrons (the Bohm type
diffusion regime) and it has a flatter energy dependence at MeV
regime energies.  The synchrotron losses of 10 TeV regime electrons
in magnetic fields of $\sqrt{<B^{2}>} > 10^{-5}$ G are faster than
the inverse Compton losses that are dominated by CMB photon
scattering.

The scale sizes of the particle distributions of the shock upstream
(both electrons and protons) for DSA is $\Delta^u \approx
\kappa(p)/u_{\rm sh}\sim 3 \times 10^{17} \cdot u_{\rm sh8}\cdot
B_{\rm \mu G}^{-1}\cdot E_{TeV}$ cm, where the r.m.s. magnetic field
$B_{\rm \mu G}$ is in $\mu$G units,  $u_{\rm sh8}$ is the shock
velocity in units of 1,000 $\kms$, and $E_{TeV}$ is the electron
energy in TeV units. In the shock upstream the width of the layer
where the highest energy electrons are stopped due to synchrotron
losses is about $\Delta^u$. In the shock downstream the width of the
ultra-relativistic electron/positron cooling layer $\Delta_{\rm
s}^d$ is about $\Delta^d \sim 6 \times 10^{21} \cdot u_{\rm
sh8}\cdot B_{\rm \mu G}^{-2}\cdot E_{TeV}^{-1} $ cm. Both widths
$\Delta^{u,d} $ of the electron regions emitting X-rays are
relatively narrow, typically below 0.3 pc for $B_{\rm \mu G} > 30$
and $E_{TeV} \gg 1$. Therefore, for large enough SNR shells of radii
$R_{\rm SNR}
>> \Delta^{u,d}$ the one dimensional approximation for the
determination of the distribution function is well justified. It is
also important that the wavelengths of magnetic fluctuations  in the
SNR shell are of the order of the gyroradii of the relativistic
protons in DSA models, and therefore that these are below
$\Delta^{u,d}$ justifying the use of a homogeneous r.m.s. field in
the losses term of the kinetic equation. We numerically calculated
the electron distribution function in the vicinity of the SNR
forward shock. The results were then used in simulations of the maps
of polarized synchrotron emission of the SNR.

\subsection{Geometry}\label{geometry}
Figure~\ref{geom} shows a 3-D sketch of the simulated SNR and its
projection along the line of sight. To simulate the images of the
SNR shell we assumed that a quarter part of a spherical forward
shock has a relativistic electron distribution $N(z, \gamma, t)$
that does not depend on the azimuthal and polar angles, but is
inhomogeneous in the radial direction with a strong peak (of width
$\Delta_{\rm s}$) at the shock position at $r =R_{\rm SNR}$. The
line of sight is along the $z$ axis.  The Stokes parameters
$\tilde{I}, \tilde{Q},\tilde{U}, \tilde{V}$ for incoherent photons
are additive, so we can integrate these over the line of sight
weighted with the distribution function of emitting particles $N(z,
\gamma, t)$ to get the intensity
\begin{equation}
\hat{S}({\bf R_{\perp}}, t,{\nu}) =\! \int dz\, d\gamma\, N(z,
\gamma, t')\,\, \hat{\tilde{S}}({\bf r},\gamma, t').{\label{Stokes}}
\end{equation}
To collect the photons reaching the observer at the same
moment $t$, we
performed an integration 
over the source depth using the retarded time $t' = t - |{\bf
r}-{\bf R_{\perp}}|/c$ as argument in ${\bf B}({\bf r}, t')$ and
$N({\bf r}, E, t')$. The integration grid has a cell size smaller
than $\Lmin$. The result is a surface density of Stokes parameters
of radiation from the volume along the line of site. The fourth
Stokes parameter V is zero in the case of an isotropic electron
velocity distribution. In order to achieve a few percent accuracy we
integrated over 8000 grid points along the line of sight. The number
of pixels in the sky projection is $100 \times 200$. The degree of
polarization was derived following \citet{gs65}.

%

 Below we present
synchrotron images simulated with a steady model distribution of
electrons accelerated by a plane shock of velocity 2,000 $\kms$
propagating in a fully ionized plasma of number density 0.03$\cmc$.
The kinetic model used to simulate the electron distribution was
described in detail by \citet{bceu00}. The magnetic field in the
far-upstream region was fixed at 3
  $\mu$G and it was assumed that the magnetic-field amplification produces a
  random field of $\sqrt{\langle B^2\rangle} = 3\times 10^{-5}$ G in the shock
  vicinity.  In the random magnetic field simulations we
  used a wavenumber range $\Kmin < k < \Kmax$, where
$\Lmin = 2\pi/\Kmax = 2\times 10^{-4}\pi \cdot D$ for $\delta = 1.0$
and  $\Lmin = 2\pi/\Kmax = 2\times 10^{-3}\pi \cdot D$ for $\delta
=2.0$, with $\Lmax = 2\pi/\Kmin = 0.2 \pi \cdot D$ for both $\delta$
values. Here $D$ is the size of a unit cubic box, as shown in
Figure~\ref{geom}. The random magnetic field in the simulated SNR
shell quarter was divided into 8 such boxes. This number of boxes
was chosen to achieve the required accuracy of the integration of
the random field along the line of sight. The field in the boxes was
simulated as a function of global SNR coordinates as it is shown in
Figure~\ref{geom} (i.e., not just locally  in each box). Note that
the field was actually simulated in a region larger than the SNR
shell and that the box sizes are larger than the sizes of the random
filamentary structures that appear.

\section{Simulated polarization maps of the X-ray synchrotron emission}\label{maps}
\begin{figure}
\centering {
 \rotatebox{0}{
{\includegraphics[height=5.6cm, width=7.5cm]{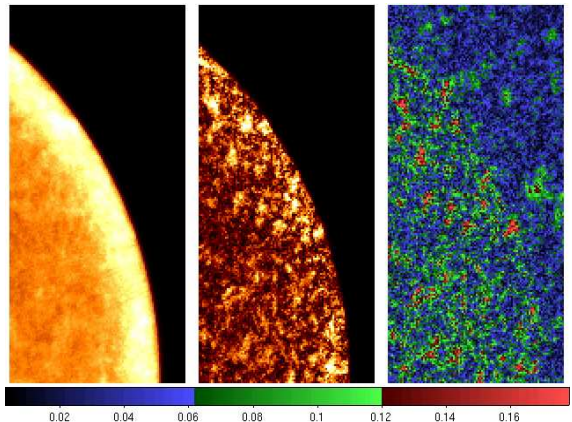} }}}\caption
{Simulated maps of polarized synchrotron
  emission in a random magnetic field at 0.5 keV.
  Intensity, $\nu^2\cdot I({\bf
  R_{\perp}},t, {\nu})$, is shown with a linear color scale in the left panel.
  The central panel  shows the product of intensity and polarization
  degree. The right panel shows the degree of polarization
  indicated by the colorbar. The stochastic magnetic field sample has $\sqrt{\langle B^2\rangle}
= 3\times 10^{-5}$ G and spectral index $\delta = 1.0$.
  }
  \label{mapK1_05}
\end{figure}

\begin{figure}
\centering {
 \rotatebox{0}{
{\includegraphics[height=5.6cm, width=7.5cm]{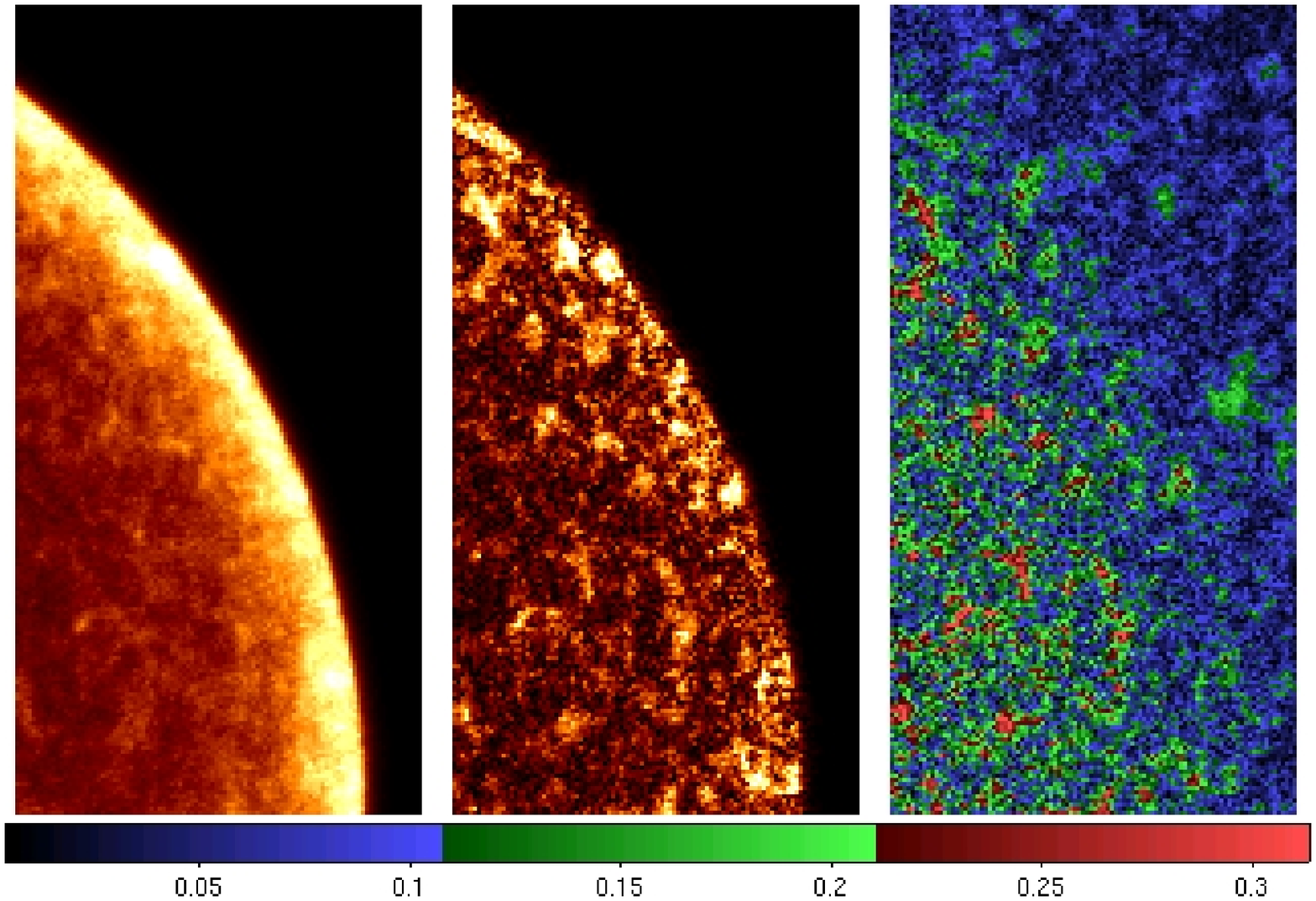} }}} \caption
{The same maps as in Figure~\ref{mapK1_05}, but at 5.0 keV.
  }
  \label{mapK1_5}
\end{figure}

\begin{figure}
\centering {
 \rotatebox{0}{
{\includegraphics[height=5.6cm, width=7.5cm]{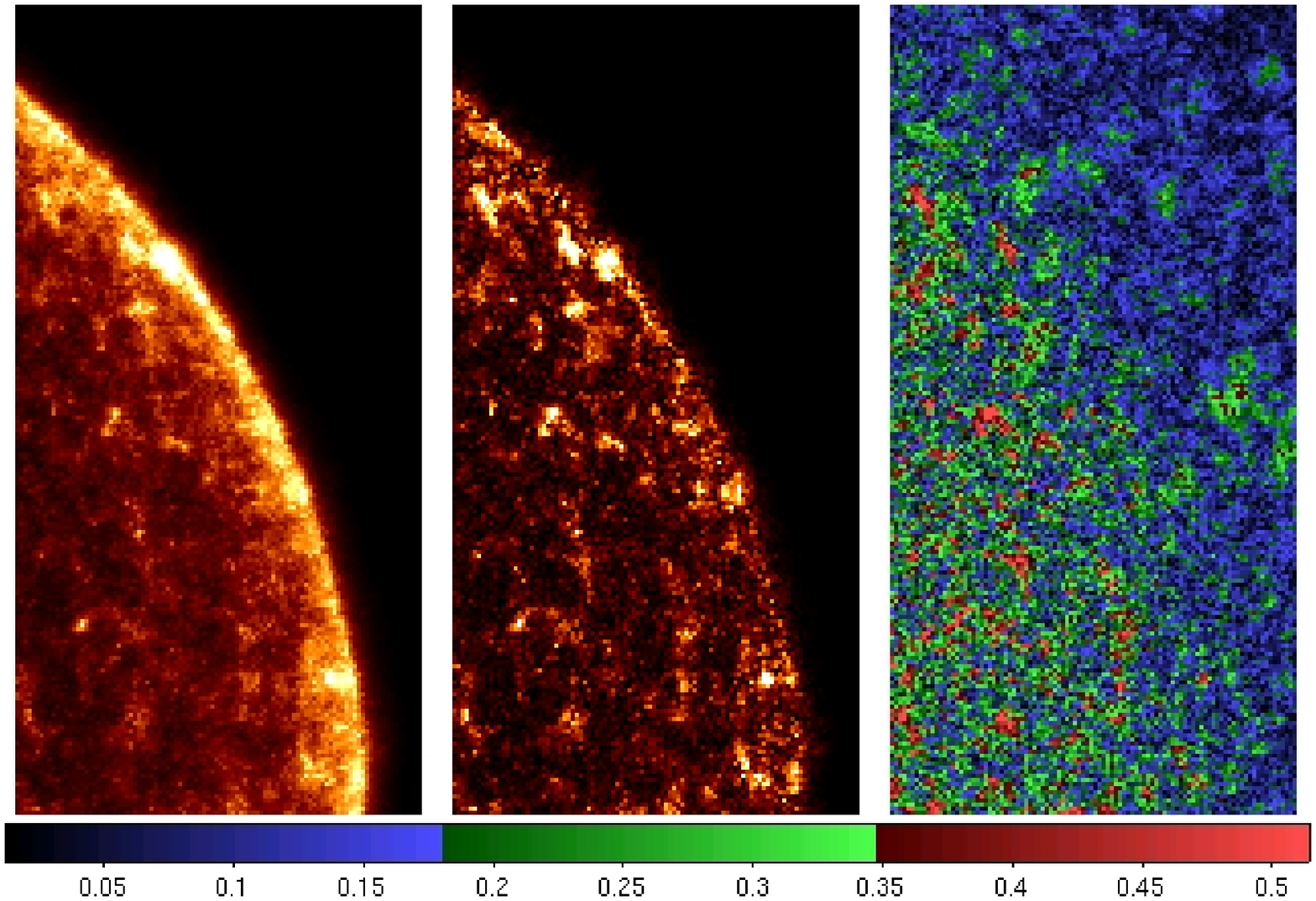} }}} \caption
{The same maps as in Figure~\ref{mapK1_05}, but at 50.0 keV. }
  \label{mapK1_50}
\end{figure}

Figures~\ref{mapK1_05}-\ref{mapK1_50} show examples of the resulting
maps at different X-ray energies. The left panels show the
synchrotron intensity, the right panels the polarization degree, and
the central panels the product of the two. The latter is a measure
of the polarized flux and is meant to illustrate that peaks in the
polarization-degree map do not necessarily correspond to peak
intensities. The images clearly demonstrate 1) the presence of
detailed structures - clumps and filaments - produced by the
stochastic field topology \citep[for details see][]{bue08} and 2)
that some of these structures emit highly polarized emission ($>
30\%$) at energies of ~5 keV and above.

\begin{figure}
\centering {
 \rotatebox{0}{
{\includegraphics[height=5.6cm, width=7.5cm]{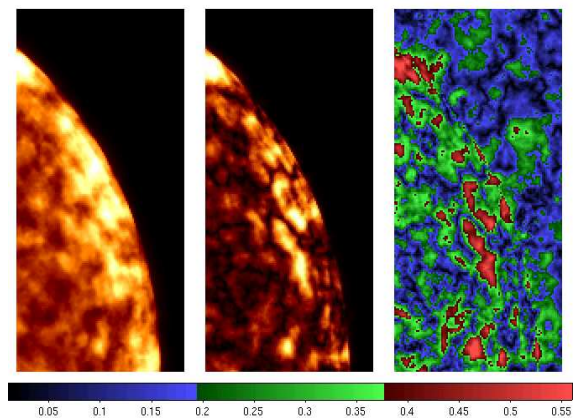} }}}\caption {
The 5.0 keV synchrotron map for a different magnetic field spectrum
in the shell. The stochastic magnetic field sample has
$\sqrt{\langle B^2\rangle} =
  3\times 10^{-5}$ G and $\delta = 2.0$.}
  \label{mapK2_5}
\end{figure}

Figure~\ref{mapK2_5} shows the 5 keV map (as in Figure~\ref{mapK1_5}),
but for a steeper spectrum of magnetic fluctuations ($\delta =
2.0$ rather than 1.0). There is a distinct difference, indicating that for
steeper spectra the size of the polarized structures is larger and
the degree of polarization of these structures is higher
(about 50\% for $\delta = 2.0$ and
$\sqrt{\langle B^2\rangle} = 3\times 10^{-5}$ G ).

Irrespective of the precise value of $\delta$, it is clear from
Figures~\ref{mapK1_05}-\ref{mapK2_5} that the degree of polarization
is higher at higher X-ray energies. The physical reason is best
illustrated in case of a power-law electron distribution (with
spectral index $\Gamma$). Namely, the degree of polarization
$\tilde{\Pi}$ and the local synchrotron emissivity $\tilde{I}({\bf
r},t, {\nu})$ have the following dependencies on $\Gamma$:
$\tilde{\Pi} \approx (\Gamma + 1)/(\Gamma + 7/3)$ (i.e. the degree
of of polarization $\tilde{\Pi}$ is increasing with $\Gamma$) and
$\tilde{I}({\bf r},t, {\nu}) \propto B^{1/2(\Gamma + 1)}$ (i.e. the
local emissivity is relatively very high for large $B$ and large
$\Gamma$) \citep[see e.g.,][]{gs65,rl79}. In the high-energy cut-off
regime the electron spectrum is typically exponential, but the
effective index $\Gamma$ is large indeed and the value increases for
electrons emitting at higher frequency $\nu$. This explains the
increase of the polarization degree with $\nu$. In addition, the
dependency of the emissivity $\tilde{I}$ on $B$ and $\Gamma$ can
lead to a highly polarized bright feature that stands out in the map
for even a single strong local field maximum. In lower-energy maps,
for which $\Gamma$ on average is smaller (i.e. well below the
cut-off regime), high polarization of a single maximum can be
smoothed or washed out by contributions from a number of weaker
field maxima integrated over the line of sight. This effect can be
seen in Figures~\ref{mapK1_05}-\ref{mapK1_50}. The high-energy maps
are 'twinkling' because of the finite life-time of the
magnetic-field amplifications. The timescale for variations in the
polarization (and the energy dependence) is similar to that of the
time variability of the intensity maps \citep[studied in \S4.1
of][]{bue08}.

\begin{figure}
\centering {
 \rotatebox{0}{
{\includegraphics[height=5.6cm, width=7.5cm]{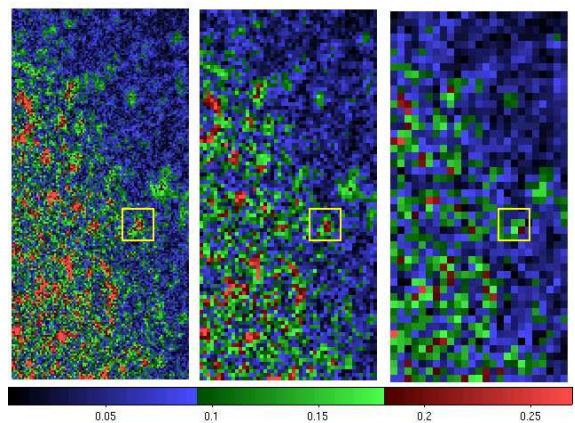} }}}\caption {
The simulated 5.0 keV synchrotron polarization maps with different
pixel sizes. The left one has a pixel size of 9\arcsec, the central
of 18\arcsec, and the right of 36\arcsec. The yellow frame of
2.6\arcmin$\times$2.6\arcmin\, indicates the
  field of view of the {\sl XPOL} polarimeter (see text). The stochastic
magnetic field sample has $\sqrt{\langle B^2\rangle} =
  3\times 10^{-5}$ G and $\delta = 1.0$. The simulated SNR shell has the radius of about 0.4$^{\circ}$.}
  \label{binned}
\end{figure}

\begin{figure}
\centering {
 \rotatebox{0}{
{\includegraphics[height=5.6cm, width=7.5cm]{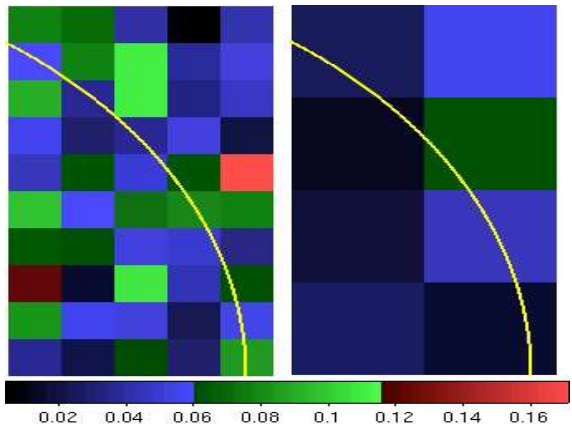} }}}\caption {
The simulated 20.0 keV synchrotron polarization maps with different
pixel sizes. The left one has a pixel size of 3\arcmin\, and the
right of 7.5\arcmin\,. The yellow line indicates the forward shock
position. The stochastic magnetic field sample has $\sqrt{\langle
B^2\rangle} =
  3\times 10^{-5}$ G and $\delta = 1.0$.
  The field was simulated in a  box larger than the SNR shell, but the regions well outside the
  forward shock are dim as it is clearly seen in the left panels in Figures~\ref{mapK1_05} - \ref{mapK2_5}.}
  \label{binned_k1}
\end{figure}

\begin{figure}
\centering {
 \rotatebox{0}{
{\includegraphics[height=5.6cm, width=7.5cm]{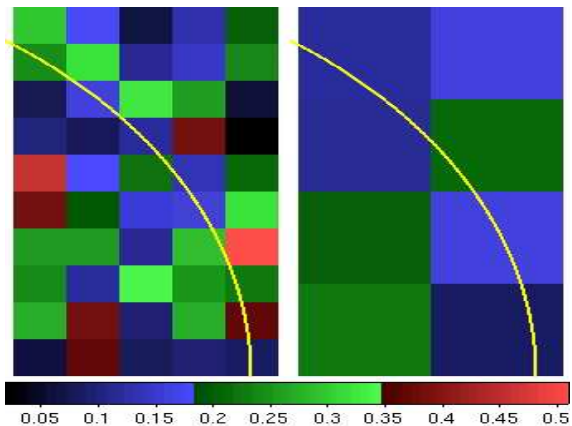} }}}\caption {
The same maps as in Figure~\ref{binned_k1}, but for magnetic
turbulence with $\delta = 2.0$. }
  \label{binned_k2}
\end{figure}

Figure~\ref{binned} illustrates the dependency of the polarization
degree on the resolution of the simulated maps (9\arcsec, 18\arcsec,
and 36\arcsec) at 5 keV. In Figures~\ref{binned_k1} and
\ref{binned_k2} the polarization maps are presented at 20 keV for
larger pixel sizes of 3\arcmin\ and 7.5\arcmin\ (close to the
INTEGRAL ISGRI pixel size) and $\delta =1$ and 2. Comparison of
Figures~\ref{binned_k1} and \ref{binned_k2} shows again the strong
dependence on the spectral index $\delta$ of the stochastic magnetic
field.

\section{Observational perspective}\label{discussion}

This work was stimulated by the fact that a number of X-ray
polarimeter instruments is being considered currently. The Imaging
X-ray Polarimetry Explorer {\sl IXPE} was proposed by
\citet{weisskopfea08} as a dedicated X-ray-polarimetry observatory
to measure the X-ray linear polarization as a function of energy,
time, and position. \citet{legereea05} is developing a Compton
polarimeter to measure polarization of hard X-rays in the 50-300 keV
energy range. A balloon-borne hard X-Ray polarimeter {\sl HX-POL} is
proposed by \citet{krawczynsk0ea08}. A Hard X-ray Telescope {\sl
HET} aboard the Energetic X-ray Imaging Survey Telescope {\sl EXIST}
\citep{grindlay09}, with a wide field of view for the coded aperture
imaging, is being designed to study the polarization at high
energies and its temporal evolution. Polarization detection of X-ray
sources as faint as 1 milliCrab is an aim of the Gravity and Extreme
Magnetism SMEX {\sl (GEMS)} mission that uses foil mirrors and Time
Projection Chamber detectors \citep{swankea08, jahodaea08}. The
missions listed above have good perspectives in this field of
research.
We address here the potentials of the {\sl XPOL} polarimeter as
proposed for XEUS \citep{costaea08} - although evolved into
International X-ray Observatory ({\sl IXO}) in the mean time - to
illustrate the observational possibilities for synchrotron X-ray
studies of SNR RXJ1713.72-3946 as a generic example.

RXJ1713.72-3946 has an extended shell of about one degree angular
diameter. A field of view of 1.5\arcmin $\times$ 1.5\arcmin\ and an
angular resolution of 5\arcsec\, was proposed by \citet{costaea08},
if {\sl XPOL} was part of the {\sl XEUS} mission. A polarimeter like
{\sl XPOL} aboard the {\sl IXO} mission
\footnote{http://ixo.gsfc.nasa.gov/science/performanceRequirements.html}
will have a somewhat larger field of view. We illustrate that case
in Figure~\ref{binned} where  the field of view of 2.6\arcmin
$\times$ 2.6\arcmin\ is shown as a yellow box in the polarization
maps simulated with pixel sizes of 9\arcsec, of 18\arcsec, and of
36\arcsec\  to illustrate the angular resolution effect.

In the case of an extended source like RXJ1713.72-3946 first of all
a wide field map of the source region is needed to specify the {\sl
XPOL} pointing. Using {\sl Chandra} archive data, we estimate the
2-10 keV flux from a 2.6\arcmin $\times$ 2.6\arcmin\ region in the
shell of RXJ1713.72-3946 to be about 4.5$\times$ 10$^{-12} \enf$.
From the minimum detectable polarization as a function of observing
time as presented in Fig. 6 of \citet{costaea08}, corrected for the
reduced effective area of {\sl IXO}$^{1}$, we estimated that a
meaningful polarization map can be constructed with {\sl XPOL}
within an exposure time of 100 ks. The polarization map of
2.6\arcmin $\times$ 2.6\arcmin\ should likely reveal detailed
highly-polarized structures with typical scales of about 10\arcsec\
as it was discovered in {\sl Chandra} images by
\citet{uchiyamaea07}. The degree of polarization will increase with
increasing X-ray energy as predicted from the modeling in this
paper. The polarization will be time variable on a few year time
scale (depending on the photon energy) as it was found by
\citet{bue08} in simulated intensity maps. Mapping of the whole
extended shell of RXJ1713.72-3946 in polarized X-rays would be
unrealistic with that set up, therefore the target must be first
identified with a wide field X-ray imager.

Another object of great interest is Cas A, that is of an angular
size comparable with the field of view of {\sl XPOL}. We estimate
that some thin peripheral polarized X-ray filaments of some ten
arcsecond scale can be studied with {\sl XPOL}, also with an
exposure of about 100 ks. In the DSA model the scale size $\Lmax$ of
the magnetic fluctuations responsible for the twinkling polarized
structures is expected to be connected to gyroradii of accelerated
protons at maximal energies. These can be roughly estimated to be
about  3 $\times10^{17} \cdot B_{\rm \mu G}^{-1}\cdot E_{100TeV}$
cm. Therefore, their angular sizes are expected to be above a few
arcseconds for SNRs within a few kiloparsec distance.

\section{Conclusions}

We have studied the {\sl polarization} of X-ray synchrotron emission
from SNRs  addressing the significant effect of magnetic-field
fluctuations on synchrotron emission in X-rays. Such magnetic
fluctuations form a natural starting point because they must be
present if diffusive shock acceleration is indeed the basic
mechanism for accelerating particles in SNRs. Like \citet{bue08} we
simulated random magnetics field to construct synchrotron emission
maps, given a smooth and steady distribution of electrons, but now
with special attention to the polarization of the resulting
emission. The simulated random magnetic fields show {\sl non-steady
localized structures with exceptionally high magnetic-field
amplitudes}. These magnetic-field concentrations dominate the
synchrotron emission - integrated along the line of sight - from
energetic $>$TeV electrons, i.e. in the cut-off regime. In terms of
a power-law electron spectrum with spectral index $\Gamma$, this can
be understood since the synchrotron emissivity $\tilde{I}({\bf r},t,
{\nu})$ is proportional to $B^{1/2(\Gamma + 1)}$ (i.e. the local
emissivity is relatively very high for large $B$ and large
$\Gamma$). The power-law approximation is only useful over a narrow
electron energy range in the cut-off regime, where the effective
spectral index $\Gamma$ is increasing with the electron energy.

Starting from the simulated magnetic-field simulations, we have
constructed maps of {\sl polarized} X-ray emission of SNR shells.
These are highly clumpy with high polarizations up to 50\%. This
characteristic of high polarization again applies to energetic
$>$TeV electrons in the cut-off regime. In terms again of a
power-law electron spectrum with spectral index $\Gamma$, this can
be understood since the degree of polarization is given by
$\tilde{\Pi} \approx (\Gamma + 1)/(\Gamma + 7/3)$ (i.e.
$\tilde{\Pi}$ is increasing with $\Gamma$).

The distinct characteristic of the modeled synchrotron emission is
its strong intermittency, directly resulting from the exceptionally
high magnetic-field amplifications randomly occurring as shown in
the simulations. Also characteristic is the increase of the
polarization degree with X-ray energy addressed in \S{\ref{maps}}.
Since this "twinkling" polarized X-ray emission of SNRs depends
strongly on the magnetic-field fluctuation spectra, it provides a
potentially sensitive diagnostic tool.

 The intermittent appearance of the polarized X-ray emission maps
of young SNR shells can be studied in detail observationally with
imagers of a few arcsecond  resolution, though even arcmin
resolution images can provide important information as it is
illustrated in Figures~\ref{binned},\ref{binned_k1},\ref{binned_k2}.
The polarized emission clumps of  arcsecond  scales are time
variable on a year or longer (depending on the observed photon
energy, magnetic field amplification factor and the plasma density
in the shell) allowing for rather  long exposures even in the hard
X-ray energy band. Hard X-ray observations in the spectral cut-off
regime are the most informative to study the magnetic fluctuation
spectra and the acceleration mechanisms of ultra-relativistic
particles.

Altogether, the modeled appearance and its time variability - on a
timescale of typically a year - resembles closely what is observed
already in X-ray images of some young supernova remnants. Observing
the predicted high polarization in clumps and filaments, however,
should probably await future instruments that are currently being
considered. Such observations will provide unique information on
magnetic fields and high-energy particle acceleration in SNRs.

\section*{Acknowledgments}
We thank the anonymous referee for careful reading of our paper and
a useful comment. Some of the calculations were performed at the
Supercomputing Centre (SCC) of the A.F.Ioffe Institute,
St.Petersburg. A.M.B. thanks R.Petre for a discussion of the {\sl
SMEX} project perspective. A.M.B. and Yu.A.U were supported in part
by RBRF grant 09-02-12080 and by the RAS Presidium Programm. SRON is
supported financially by NWO, the Netherlands Organisation for
Scientific Research.

\bibliographystyle{mn2e}
\bibliography{xpol1r}

\end{document}